\documentclass[sigconf]{acmart}

\usepackage[flushleft]{threeparttable}
\usepackage{graphicx}
\usepackage[disable]{todonotes}
\usepackage{url}
\usepackage{amsmath}
\usepackage{subfig}
\usepackage{tabularx}
\usepackage{multirow}
\usepackage{algorithm}
\usepackage[noend]{algpseudocode}
\usepackage{enumitem}
\usepackage[multiple,flushmargin]{footmisc}

\usepackage{amssymb}
\usepackage{pifont}
\newcommand{\cmark}{\ding{51}}%

\graphicspath{{images/}}
\usepackage{xcolor}

\newcommand{\Note}[2]{}
\newcommand{\SideNote}[2]{} 
\renewcommand{\Note}[2]{\todo[color=#1,size=\small, inline=true]{#2}} 
\renewcommand{\SideNote}[2]{\todo[color=#1,size=\small]{#2}} %

\newenvironment{definition}[1][Definition]{\begin{trivlist}
\item[\hskip \labelsep {\bfseries #1}]}{\end{trivlist}}

\begin{document}

\setlength{\abovedisplayskip}{5pt}
\setlength{\belowdisplayskip}{5pt}

\acmPrice{15.00}

\settopmatter{printacmref=true}
\fancyhead{}




\title{Context Attentive Document Ranking and Query Suggestion}

\author{Wasi Uddin Ahmad}
\affiliation{%
  \institution{University of California, Los Angeles}
  \city{Los Angeles}
  \state{CA}
  \postcode{90095}
}
\email{wasiahmad@ucla.edu}

\author{Kai-Wei Chang}
\affiliation{%
  \institution{University of California, Los Angeles}
    \city{Los Angeles}
  \state{CA}
  \postcode{90095}
}
\email{kwchang.cs@ucla.edu}

\author{Hongning Wang}
\affiliation{%
  \institution{University of Virginia}
  \city{Charlottesville}
  \state{VA}
  \postcode{22904}
}
\email{hw5x@virginia.edu}

\begin{abstract}
We present a context-aware neural ranking model
to exploit users' on-task search activities and enhance retrieval performance. In particular, a two-level hierarchical recurrent neural network is introduced to learn search context representation of individual queries, search tasks, and corresponding dependency structure by jointly optimizing two companion retrieval tasks: document ranking and query suggestion. To identify variable dependency structure between search context and users' ongoing search activities, attention at both levels of recurrent states are introduced. Extensive experiment comparisons against a rich set of baseline methods and an in-depth ablation analysis confirm the value of our proposed approach for modeling search context buried in search tasks.
\end{abstract}




\maketitle

\section{Introduction}

The scope and complexity of users' information need never get simpler \cite{agichtein2012search}. To fulfill a complex need, e.g., job hunting, users issue a series of queries, exam and click search results from multiple sites. 
Such search behavior is usually referred to as search tasks \cite{jones2008beyond,wang2013learning} or sessions, which are characterized by rich types of user-system interactions, implicit feedback, and temporal dependency among the search activities.
Various studies have shown that exploring users' on-task search activities to enrich retrieval models is effective for improving retrieval performance, especially when users' intent is ambiguous. For example, 
through a large-scale analysis of search engine logs, Bennett et al. \cite{bennett2012modeling} showed that a user's short-term search history becomes more important as the search session progresses. White et al. \cite{white2013enhancing} reported the use of users' on-task behavior yielded promising gains in retrieval performance in the Microsoft Bing search engine.

However, limited by the devised form of representation for search context, most existing solutions model users' on-task behavior in an ad-hoc manner. Typically, keywords or statistical features are extracted from previous clicks or queries \cite{shen2005context,bennett2012modeling,white2013enhancing}, or manually crafted rules are introduced to characterize the changes in a search sequence \cite{guan2013utilizing,xiang2010context}. Those algorithms' exploration of contextual information is thus subjected by the capacity of their employed representation, which can hardly be exhaustive nor optimal for the retrieval tasks of interest. For example, keyword-based methods suffer from vocabulary gap, and statistical features become unreliable with sparse observations.
Even if a rich set of contextual features can be provided beforehand, the dependency structure has to be imposed a priori, e.g., either to use the immediate one preceding query or all queries in a task to calculate the feature values. This cannot capture variable range dependency within a user's sequential search activities.

Moreover, during a search task users have to get involved in multiple retrieval tasks.
For instance, to perform a search task, not only does a user need to respond to the system's returned search results (e.g., examine or click), but also to the suggested queries (e.g., accept or revise the suggestions).
Arguably, when concurrently performing these retrieval tasks, users are motivated by the same underlying intent, and therefore their search activities are interrelated across the companion retrieval tasks. This dependency reveals fine-grained search context beyond the content of submitted search queries and clicked documents. For example, 
if a user skipped a top-ranked document, the suggestion for next query should be less related to such documents.
Inspired by these scenarios, recent works \cite{liu2015representation,ahmad2018multitask, salehi2018multitask,huang2018improving,nishida2018retrieve} have proposed to jointly model multiple types of user search activities.
These solutions focus mostly on using an auxiliary task to assist the target task with two objectives: (1) leveraging large amount of cross-task data, and (2) benefiting from a regularization effect that leads to more useful representations.
However, none of these multi-task retrieval solutions model the sequential dependency across different retrieval tasks. This inevitably limits their ability in exploiting information buried in a user's search sequence.

To address the aforementioned challenges in modeling users' on-task search behaviors, we present a context-aware neural retrieval solution, \emph{Context Attentive document-Ranking and query-Suggestion} (CARS). 
Given a query and the user's past search activities (e.g., his/her issued queries and clicks) in the same search task, CARS encodes them into search context representations.
Based on the learnt representations, CARS then predicts the ranking of documents for the given query and in turn suggests the next query.
To encode search context, we employ a two-level hierarchical recurrent neural network.
At the lower level, given queries and documents as a sequence of words, we encode them using bidirectional recurrent neural networks;
and at the upper level, we introduce another layer of recurrent states on top of the embedding vectors of queries and documents to represent task-level search context. 
Each observed action of query reformulation or result click contributes to the update of task-level recurrent states, which thus serve as a learned summary of past search activities, providing relevant information for predicting document ranking and next query. 
To identify variable dependency structure between search context and ongoing user search activities, we apply attention mechanism at both levels of recurrent states. 
This endows CARS to model the development of users' search intent in the course of search tasks.

To learn search context representation and corresponding dependency structure, CARS jointly optimize for two companion retrieval tasks, i.e., document ranking and query suggestion.
CARS models the relatedness between these two tasks via a regularized multi-task learning approach \cite{evgeniou2004regularized}.
We evaluate CARS on the AOL search log, the largest publicly available search engine log with both authentic user query and click information. We compared our model with a rich set of baseline algorithms (both classical and neural IR models), which model users on-task behavior differently for document ranking and query suggestion. Extensive experiment comparisons and significant improvements over the baselines confirm the value of modeling search context buried in search tasks.

\section{Related Works}
\label{sec:rel-work}

Context information embedded in a search task has shown to be useful for modeling user search intent \cite{bennett2012modeling,jones2008beyond,liao2012evaluating}. A rich body of research has explored different forms of context and search activities and built predictive models to improve retrieval performance. The related works can be roughly categorized as data-driven v.s., model-driven solutions for task-based retrieval. 

Data-driven solutions focus on deriving contextual features from users' search activities to characterize their search intent. Shen et al. \cite{shen2005context} extract keywords from users' past queries and clicked documents in a search session to re-rank document for future queries. White et al. \cite{white2010predicting,white2013enhancing} develop a rich set of statistical features to quantify context information from users' on-task search behavior. Xiang et al. \cite{xiang2010context} craft a collection of rules to characterize the search context, e.g., specialization v.s., generalization, so as to extract features by the rules. As we discussed before, data-driven solutions are confined by their employed form of context representation, e.g., keywords or manually crafted rules, which is hardly generalizable or optimal with respect to different retrieval tasks.

Model-driven solutions build predictive models about users' search intent or future search behavior. 
Cao et al. \cite{cao2009towards} model the development of users' search intent in search sessions with a variable length Hidden Markov Model, and utilize the inferred search intent for document ranking and query suggestion. 
Reinforcement learning is utilized to model user-system interactions in search tasks \cite{guan2013utilizing,luo2014win}. Syntactic changes between consecutive queries and the relationship between query changes and retrieved documents, are modeled to improve retrieval results. However, the predefined model space (e.g., add/remove query terms) and state transition structure (e.g., first-order Markov chain) forbid this type of solutions from learning rich interaction between users and a system.

Encouraged by the recent success of neural network based retrieval solutions \cite{huang2013learning,lu2013deep,guo2016deep,jaech2017match,borisov2018click}, various models have been developed to optimize session-based retrieval. Mitra et al. \cite{mitra2015exploring} studies session context with a distributed representation of queries and reformulations and uses the learned embeddings to improve query prediction.  \cite{sordoni2015hierarchical,jiang2018rin,huang2018improving,wu2018query} exploited hierarchical neural architectures to model a sequence of queries in the same search session. 
Recently, Chen et al. \cite{chen2018attention} propose a hierarchical attention based structure to capture session- and user-level search behavior. However, these neural models focus on learning search context representation from single retrieval tasks, e.g., document ranking or query suggestion, and therefore cannot utilize the reinforcement between different retrieval tasks. In addition, most solutions for search task based representation learning do not differentiate the influence from different actions in a sequence. For example, clicks from a nearly duplicated query to the current query discloses more information about a user's current focus than those not similar to the current query, although that nearly duplicated query might be submitted long time ago. Recognizing such variable length dependency is crucial for modeling the search context and thus inferring users' information need. 

Multi-task learning has been explored in information retrieval studies \cite{liu2015representation,huang2018improving, salehi2018multitask,nishida2018retrieve}. The basic idea is to use one learning task as  regularization for another task. For example, 
Liu et al. \cite{liu2015representation} proposed a multi-task deep neural approach to combine query classification and document ranking, and showed improvement on both tasks. 
Huang et al. \cite{huang2018improving} coupled context-aware ranking and entity recommendation to enhance entity suggestion for web search.
Similarly, Salehi et al. \cite{salehi2018multitask} adopted semantic categorization of the query terms to improve query segmentation.
From a different angle, Ahmad et al. \cite{ahmad2018multitask} proposed to  train a document ranker and a query recommender jointly over a sequence of queries in a session.
However, none of the existing multi-task solutions paid attention to the dependency structure embedded in a search task, which characterizes users' search intent.
In this work, we explicitly model the dependency between users' in-session query and click sequence by learning context attentive representations, which mutually enhance document ranking and query suggestion.

\section{A Context Attentive  Ranking and Suggestion Model}


\begin{figure}[t]
\centering
\includegraphics[width=1.0\linewidth]{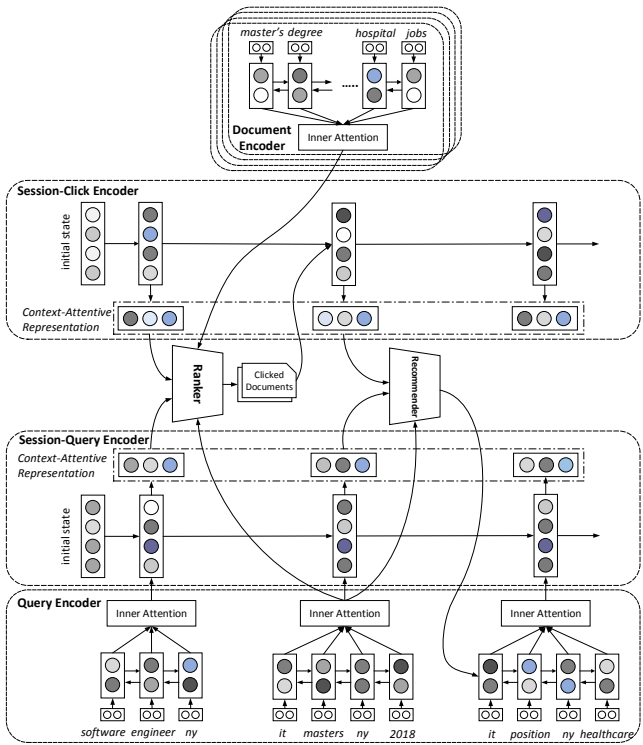}
\vspace{-4mm}
\caption{System architecture of the Context Attentive document Ranking and query Suggestion (CARS) model. Our key novelty is to encode information in search actions and on-task search context into context-attentive representations to facilitate document ranking and query suggestion tasks.}
\label{model}
\vspace{-3mm}
\end{figure}


\subsection{Problem Statement}
In a search task, a user keeps formulating queries, examining and clicking search results until his/her information need is satisfied \cite{jones2008beyond,wang2013learning}. A user's search activities in the same task, e.g., query reformulation and result clicks, often exhibit strong inter-dependency, which provides rich context information for systems to improve their retrieval performance \cite{liao2012evaluating,guan2013utilizing,luo2014win,white2013enhancing}. However, as the users' information need and behavior pattern vary significantly from task to task, modeling the search context and its use in specific retrieval problems is the key to unleash its vast potential.

Assuming a user submits a query ``it masters ny 2018'',
a common interpretation of it could be that the user is looking for the latest IT master's degree programs in New York.
However, if we knew that the user just followed a suggested query ``software engineer ny'' several queries before, it becomes evident that the user is actually looking for a software engineer position in New York, and he/she has a master's degree in IT. Hence, the search engine should promote job listings in the region that match the user's qualification and make more specific query suggestions (e.g., target at different industries). 
As the task progresses, if the user's next clicked results reflect his/her interest in healthcare industry, the system can further customize the search results and specialize its suggested queries (e.g., suggest names of particular companies in healthcare industry). 
By inferring the user intent behind each query reformulation and result click regarding the context of his/her immediate interaction history, a search engine can rapidly improve its service quality as the search task progresses.

In this work, we propose a framework to explicitly model search context using representation learning to improve both document ranking and query suggestion in a search task.
To the best of our knowledge, our proposed \emph{Context Attentive document Ranking and query Suggestion} (CARS) model is of its first kind where both a user's query and click sequences from an ongoing search task are utilized to learn the search context representation and optimize two distinct retrieval tasks jointly.

In a nutshell, CARS maintains a two-level hierarchical recurrent neural network (RNN) structure for learning in-task search context representation. 
The system architecture of CARS is illustrated in Figure \ref{model}.
At the lower level, RNN-based query and document encoders encapsulate information in a user's query formulation and click actions into continuous embedding vectors; and at the upper level, another set of RNN-based query- and document-session encoders take the embeddings of each search action as input and summarize past on-task search context on the fly. Then, the learned representations from both levels are utilized to rank documents under the current query and suggest the next query. 

Before we zoom into the details of each component, we first specify the definitions of several important concepts and the notations. 
We represent a user's search history as a sequence of queries $\mathcal{Q}=\{q_{1},q_{2},\dots,q_{N}\}$, where each query $q_{i}$ is associated with a timestamp $t_{i}$ when the query is submitted and the corresponding list of returned documents, $\mathcal{D}_i=\{d_{i,1},d_{i,2},\dots,d_{i,M}\}$. Each query $q_{i}$ is represented as the original text string that users submitted to the search engine, and $\mathcal{Q}$ is ordered according to query timestamp $t_{i}$. Each returned document $d_{i,m}$ has two attributes: its text content and click timestamp $c_{i,m}$ ($c_{i,m}=0$, if it was not clicked). In general, user clicks serve as a good proxy of relevance feedback \cite{joachims2007evaluating,joachims2017accurately}, and they serve as the training signals for our document ranker. In this work, we follow Wang et al. \cite{wang2013learning}'s definition of search tasks: 

\begin{definition} \textbf{(Search Task)} Given a user's search history $\mathcal{Q}$, a search task $\mathcal{T}_{k}$ is a maximum subset of queries in $\mathcal{Q}$, such that all the queries in $\mathcal{T}_{k}$ correspond to a particular information need.
\end{definition}

As a result, $\{\mathcal{T}_{k}\}_{k=1}^K$ is a set of disjoint partitions of a user's search history $\mathcal{Q}$: $\forall j\ne k$, $\mathcal{T}_{j}\cap \mathcal{T}_{k}=\emptyset$ and $\mathcal{Q}=\bigcup_k \mathcal{T}_{k}$. A related concept in IR literature is search session \cite{jones2008beyond}, which is usually defined by the inactive time between two consecutive search queries. Some past research assumes each search session can associate with only one particular information need, and thus they treat a session as a task \cite{luo2014win,guan2013utilizing}. This further introduces the compounding concepts of in-session task \cite{liao2012evaluating} and across-session task \cite{wang2013learning}. CARS can be readily applied to these different types of task (or session), as long as it follows our definition above. In this work, we will not differentiate between these different realizations of search tasks, but take it as the input of our algorithm. When no ambiguity is introduced, we will use the terminology ``search task'' and ``search session'' \emph{interchangeably} in this paper. In addition, without further specification, we use $W_{:}$ and $b_{:}$ to represent a trainable weight matrix and vector, respectively as our model parameters.

\subsection{Learning Search Context Representations}
\label{sec_representation}
CARS models users' search intent buried in search tasks by jointly learning from retrieval tasks of document ranking and query suggestion.
Formally, we consider document ranking as learning a candidate document's relevance to a user's current query and search context, and query suggestion as learning the most likely query that follows the current query and search context. 
We treat queries and documents as variable length word sequences, and a search task as a sequence of queries and their result clicks.
The key in both learning tasks is the representation of search actions and search context, and the dependency structure among them.

To this end, we employ hierarchical recurrent neural networks where the lower-level networks learn the query and document representations separately and the upper-level networks model the variable length dependency structure in the search context.

\noindent\textbf{$\bullet$ Lower-level Query Document Embedding.}
The lower-level recurrent network creates a fixed-length vector to represent a variable length word sequence (e.g., query, document). CARS employs two networks with the same architecture to encode queries and documents \emph{separately}, so as to capture their heterogeneity. In essence, given a sequence of $T$ words ($w_1, \ldots , w_T$), the network first embeds the word $w_t$ into a $l_{w}$-dimensional vector $x_t$ using a pre-trained word embedding \cite{pennington2014glove}. 
Then, a bidirectional recurrent neural network (BiLSTM)  \cite{schuster1997bidirectional} with an inner-attention mechanism~\cite{liu2016learning} is used to encode the word sequence into a fixed-length vector.

Specifically, an LSTM \cite{hochreiter1997long} encodes an input sequence by sequentially updating a hidden state. 
At each step $t$, given an input word vector $x_t$ and the previous hidden state $h_{t-1}$, the hidden state is updated by $h_t = LSTM(h_{t-1},x_t)$.\footnote{We follow \cite{hochreiter1997long} to use a shorthand in representing the LSTM cell, and the detailed update rules can be found in that paper.} To better capture information presented in a word sequence, we use a BiLSTM (one forward and one backward LSTM) to encode the sequence from both directions. 
The BiLSTM forms a sequence of $T$ hidden representations, 
\begin{equation}
\label{eq:h}
H = [h_1, \ldots , h_T], \quad H \in \mathbb{R}^{2l_h\times T}
\end{equation}
by concatenating the hidden states generated by the two LSTM models, where $l_h$ is the dimension of the forward and backward LSTM hidden unit. To recognize the topical importance of each word in a given input sequence, e.g., focus of a query, we apply inner-attention to form a fixed-length sequence representation $\pi$ from the variable length sequence representation $H$,
\begin{equation}
\label{eq:inner_attn}
    \pi = H \alpha^h, \quad
    \alpha^h = \mbox{softmax}\left(W^\alpha_1 \tanh(W^\alpha_2 H + b^\alpha_1) + b^\alpha_2\right),
\end{equation}
where $\alpha^h \in \mathbb{R}^T$ is the attention vector, $\tanh(\cdot)$ is an element-wise tangent function on the input matrix, and $W^\alpha_{1}, W^\alpha_{2}, b^\alpha_{1}$ and $b^\alpha_{2}$ are the parameters of a two-layer perceptron to estimate the attention vector.
The attention vector assigns weight for each individual word in the sequence, such that informative words would play a more important role in the final sequence representation $\pi$.

When no ambiguity is invoked, we will refer to $q_i$ and $d_{i,m}$ as the sequence representations learnt for the $i$-th query and the corresponding $m$-th candidate document.

\noindent\textbf{$\bullet$ Upper-level Task Embedding.}
Within a search task, a user submits a sequence of queries, examines the returned documents, and clicks a few of them when found relevant.
To encode the search context of an on-going task, we use a pair of recurrent neural networks that operate on top of the query and click representations learnt from the lower level networks, and refer to them as session-query encoder and session-click encoder respectively. 

Query reformulation chain in a search task carries important contextual information about a user's search intent \cite{guan2013utilizing,luo2014win}. To represent search context in a query chain, we use an LSTM as the session-query encoder. This encoder takes a sequence of learned query representations till query $q_i$ as input and computes the corresponding recurrent states by $s_i^q = LSTM(s_{i-1}^q, q_i)$, where $s_i^q \in \mathbb{R}^{l_q}$ is the session recurrent state at the $i$-th query and $l_q$ is the dimension of this LSTM's hidden unit. 

A user's click sequence in a search task also contributes to its search context. But research shows that user clicks reflect their search intent from a different perspective than query reformulation chain does, and also these two types of feedback introduce distinct biases and variances in different retrieval tasks \cite{joachims2007evaluating}. 
We employ a separate task-level LSTM for the clicked documents, which we refer to as the session-click encoder.
Assume documents $\{d_{c_1}, d_{c_2}, \ldots d_{c_{N_i}}\} \subset \cup_{t=1,2,\ldots,i-1} \mathcal{D}_t$ are the clicked documents in the current search task $\mathcal{T}_k$ before query $q_i$ is submitted (according to their click  and query timestamps). The session-click encoder sequentially visits each clicked document and at the $n$-th clicked document $d_{c_n}$, the recurrent state of this LSTM is updated by $s_{n}^c = LSTM(s_{n-1}^c, d_{c_n})$, where $s_{n}^c\in \mathbb{R}^{l_c}$ and $l_c$ is the dimension of this LSTM's hidden unit.

Not all the clicked documents are equally useful to construct the search context \cite{joachims2007evaluating}, and they may depend on each other to collectively present a complete user information need. Hence, we employ the inner-attention used in Eq \eqref{eq:inner_attn} over the learned click recurrent states to recognize the importance of each different clicked document and learn their composition in an ongoing search task. 

\noindent\textbf{$\bullet$ Context Attentive Representations.}
In recurrent neural networks, it is typical to use the last hidden state as a summary of the whole sequence. However, in the scenario of task-based retrieval, the immediate past search action is not necessarily the most important to model search context \cite{bennett2012modeling}. But it is also difficult to pre-define the dependency structure. It is preferred to learn the dependency structure from a user's past interactions in the same task. 

To this end, CARS learns to represent the search context till current query $q_i$ by applying attention \cite{chen2018attention} over the whole search sequence, which accounts for the informativeness of each past search action regarding the search context and $q_i$.
To enhance search context representation, the session query recurrences are refined as follows:
\begin{equation} 
\label{eq:session_attn}
s_i^{att,q}   = \sum_{j=1}^{i-1} \alpha_j^q
    s_j^q, \quad 
    \alpha_j^q = \frac{\exp(q_i^\top W^e s_j^q)}{\sum_{k=1}^{i-1} \exp(q_i^\top W^e s_k^q)},
\end{equation}
where 
$\alpha_j^q$ is the attention weight computed against the current query representation $q_i$, session query recurrence $s_j^q$, and a learnt attention weight matrix $W^e$.
The attentive vector $s_{i}^{att,q}$ integrates the contribution of the previous in-task queries and guides the generation of current query $q_i$.

Similarly, we use this attention mechanism between $q_i$ and click recurrence states $[s_{1}^c, \ldots, s_{N_i}^c]$ to form $s_{i}^{att,c}$, which represents the document content explored by the user previously in the same task before $q_i$. 
To combine potentially complementary information from these two task-level summary vectors, we concatenate them to form our search context attentive representation, $s_i^{att} = [s_{i}^{att,q}, s_{i}^{att,c}]$ and
$s_i^{att} \in \mathbb{R}^{l_q+l_c}$. It is then used in the document ranking and query suggestion tasks.\footnote{We compute individual attention and in turn attentive vector for the document ranking ($\alpha_{j,r}$; $s_i^{att,r}$) and query suggestion ($\alpha_{j,s}$; $s_i^{att,s}$) tasks.} 
We should note that the attention applied over the past search activities recognizes their contributions in representing the search context up to the current search action, but not to a particular retrieval purpose, e.g., document ranking or query suggestion. We will discuss how to optimize these task-level representations with respect to specific retrieval tasks next.

\subsection{Joint Learning of Ranking and Suggestion}
In the following, we describe how we optimize the model parameters to learn effective search context representations.

\noindent\textbf{$\bullet$ Document Ranking.}
The goal of a document ranker is to rank the most relevant documents to the input query on top. As we do not have explicit relevance feedback from users, we use their clicks as relevance labels. To simplify the model, we appeal to the pointwise learning to rank scheme, where a ranker is designed to predict whether a document will be clicked under a given query. The documents are then ranked by the predicted click probabilities. 
In CARS, the click prediction for the $m$-th document under query $q_i$ is based on the document vector $d_{i,m}$ (see Section \ref{sec_representation}) and a composed vector $u_i$ generated by the current query vector $q_i$ and the search context attentive vector $s_i^{att}$,
\begin{equation}
\label{eq:u}
    u_i = W^u_1 s_i^{att} + W^u_2 q_i + b^u,
\end{equation}
where $W^u_1$, $W^u_2$, $b^u$ are parameters of our ranker. Albeit user clicks are known to be biased \cite{joachims2007evaluating}, empirical studies also show promising results \cite{huang2013learning}. We leave more advanced click modeling and learning to rank approaches as our future work. 

Various models can be employed here to predict click based on these two vectors.
Following \cite{hu2014convolutional,mitra2017learning}, we first create an extended matching vector to capture the similarity between $d_{i,m}$ and $u_i$, as $[d_{i,m}, u_i, (d_{i,m} - q_i), (d_{i,m} \odot q_i)]$ where $\odot$ denotes element-wise multiplication. Then we feed this matching vector to a three-layer batch-normalized maxout network 
\cite{goodfellow2013maxout} to predict the click probability $P(c_{i,m}|q_i, d_{i,m}, s_i^{att})$, denoted as $o_{i,m}$.

\noindent\textbf{$\bullet$ Query Suggestion.}
The query suggestion component (a.k.a. recommender) takes current query and search context as input to predict the next query for a user as $P(q_{i+1}|q_i, s^{att}_{i})$, which can be decomposed into a series of word-level predictions,
\begin{equation*}
P(q_{i+1}|q_i, s^{att}_{i}) = \prod\nolimits_{t=1}^{|q_{i+1}|} P(w_{t}|w_{1:t-1}, q_{i},  s^{att}_{i}).
\end{equation*}
This can be readily estimated by the decoder in a sequence to sequence network \cite{sutskever2014sequence}.

We use the search context attentive vector to initialize the hidden state $h_0^{dec}$ in the decoder by $h_0^{dec} = \tanh(W^{h_0} s^{att}_{i} + b^{h_0})$, 
where $W^{h_0}\in \mathbb{R}^{l_h \times (l_q + l_c)}$ and $b^{h_0}\in \mathbb{R}^{l_h}$ are the decoder parameters. The recurrence is computed by: $h_t^{dec} = LSTM(h_{t-1}^{dec}, w_{t-1})$, where $w_{t-1}$ is the previously generated word.
In standard use of an LSTM-based sequence decoder, the output sequence is generated by a recurrently computed latent state $h_{t-1}^{dec}$ and sampling the words accordingly. This, unfortunately, cannot carry over the search context in query word sequence generation, as the context is only used to initialize the decoder. To enhance the influence of search context in our query suggestion, we apply attention based on the search context $s^{att}_{i}$ and current query $q_i$ in the decoding process.

During a web search, users often reformulate their query by modifying a few words from their last query. 
For example, more than 39\% users repeat at least one term from their immediate previous queries \cite{jiang2014learning} and an average of 62\% terms in a query are retained from their previous queries \cite{sloan2015term}.
Motivated by this, we predict the $t$-th word in the next query $q_{i+1}$ based on a constructed attention vector $a_{i,t}$ that encodes the query terms in the current query $q_i$ with respect to the latent state of decoder at the $t$-th generated word: $a_{i,t} = \sum_{k=1}^{|q_i|} \alpha^q_{tk} h_{k}^{enc}$, where $h_{k}^{enc}$ is the $k$-th column of $H$ when encoding $q_i$ (defined in Eq \eqref{eq:h}).
The normalized attention weight $\alpha^q_{tk}$ is learned using a bilinear function, 
\begin{equation}
\label{eq:attn_sgs}
    \alpha^q_{tk} = \frac{\exp((h_t^{dec})^\top W^{\alpha^q} h_k^{enc})}  {\sum\nolimits_{j=1}^{|q_i|}   \exp((h_t^{dec})^\top W^{\alpha^q} h_j^{enc})},
\end{equation}
where $W^{\alpha^q}$ is the parameter matrix to be learned.

We concatenate the attention vector $a_{i,t}$ for current query $q_i$ with $h_t^{dec}$, combine it with the search context vector $s_i^{att}$ by 
\begin{equation}
    \label{eq:nu}\nu_{i,t} = W^\nu_1 s_i^{att}+ W^\nu_2[h_t^{dec}, c_{i,t}],
\end{equation}
and generate the next word $w_t$ in the suggested query $q_{i+1}$ based on the
 following probability distribution over the vocabulary $V$,
\begin{equation}
\label{eq:prob_dist}
    P(w_t|w_{1:t-1}, q_{i},  s^{att}_{i}) = \text{softmax}(W^{gen} \nu_{i,t} + b^{gen}),
\end{equation}
where $W^{gen} \in \mathbb{R}^ {|V| \times (l_q+l_c)}$ and $b^{gen} \in \mathbb{R}^{|V|}$ are the corresponding decoder parameters.

However, the search space for this decoding problem is exponentially large, as every combination of words in the vocabulary can be a candidate query. We follow the standard greedy decoding algorithm 
to generate the next query. Specifically, the best prefix $w_{1:t}$ up to length $t$ is chosen iteratively and extended by sampling the most probable word according to the distribution in Eq \eqref{eq:prob_dist}.
The process ends when we obtain a well-formed query containing the unique end-of-query token.

\noindent\textbf{$\bullet$ Optimizing the Representations via Multi-task Learning.}
To better couple the document ranking and query suggestion tasks for learning the search context representations, we adopt the regularization based multi-task learning technique \cite{evgeniou2004regularized} and decompose $W^u_1$ (defined in Eq \eqref{eq:u}) and $W^\nu_1$ (defined in Eq \eqref{eq:nu}) parameter matrices into $W^u_1 = W^{share} + W^{rank}$ and $W^\nu_1 = W^{share}+ W^{recom}$, where $W^u_1\in \mathbb{R}^{l_h \times (l_q + l_c)}$ and $W^\nu_1\in \mathbb{R}^{l_h \times (l_q + l_c)}$. Here, $W^{share}$ is shared between the two tasks, while $W^{rank}$ and $W^{recom}$ are kept private to the corresponding learning tasks. We choose to impose this structure to couple the two learning tasks, otherwise they would have full degree of freedom to over fit their own observations rather than collaboratively contribute to the shared search context representation learning. $W^{share}$ is thus expected to capture the homogeneity in the search context's effect in these two tasks, and $W^{rank}$ and $W^{recom}$ are to capture task homogeneity from task data accordingly.    

To estimate the model parameters in CARS, we minimize regularized negative log-likelihoods of the document ranking and query suggestion tasks,
\begin{equation*}
    \mathcal{L}_{R1} + \mathcal{L}_{R2} + \frac{1}{N} \sum\nolimits_{k=1}^N \sum\nolimits_{q_i \in \mathcal{T}_k} \Big( \mathcal{L}_{\text{ranker}}(q_i) + \mathcal{L}_{\text{recom.}}(q_{i};q_{1:i}) \Big),
\end{equation*} 
where $N$ is the number of search tasks in the training set. 
$\mathcal{L}_{\text{ranker}}(q_i)$ is the negative log-likelihood with respect to the predicted clicks under query $q_i$:
\begin{equation*}
\begin{split}
\mathcal{L}_{\text{ranker}}(q_i) = & -\frac{1}{m} \sum\nolimits_m \big(\mathcal{I}(c_{i,m}>0) \log\ o_{i,m} \\ 
 & + \mathcal{I}(c_{i,m}=0) \log(1 - o_{i,m})\big),
\end{split}
\end{equation*}
where $c_{i,m}$ and $o_{i,m}$ represent the observed user clicks and predicted click probability for the $m$-th candidate document for query $q_i$. $\mathcal{L}_{\text{recom.}}$ is the negative log-likelihood of generating query $i$ based on all previous queries and clicks in the task $\mathcal{T}_k$:
\begin{equation*}
\begin{split}
\mathcal{L}_{\text{recom.}}(q_i) = & -\sum\nolimits_{t=1}^{|q_i|} \log P(w_t | w_{1:t-1}, q_{1:i-1}, d),
\end{split}
\end{equation*}
where $d\in \{d_{j:k}|c_{j:k}=1\},j\in\{1,\ldots,i-1\}$ and $k\in\{1,\ldots,m\}$.
To avoid overfitting and prevent the predicted word distributions being highly skewed, we apply two forms of regularization.
First, we regularize the shared and private parameters $W^{share}$, $W^{rank}$ and $W^{recom}$ by $$\mathcal{L}_{R1} = \lambda_1 \|W^{share}\|_2 + \lambda_2 (\|W^{rank}\|_2 + \|W^{recom}\|_2).$$
And, we add the negative entropy regularization $$\mathcal{L}_{R2} = \lambda_3 \sum\nolimits_{w \in V} P(w|q_{1:i-1}, w_{1:t-1}) \log P(w|q_{1:i-1}, w_{1:t-1})$$ as suggested in \cite{ahmad2018multitask} to smooth the predicted word distribution.

\section{Experiments and Results}


\begin{table}[t]
\centering
\caption{ Statistics of the constructed evaluation dataset based on AOL search log.}
\vspace{-2mm}
\begin{tabular}{l|c|c|c}
\hline
Dataset Split & Train & Validation & Test \\
\hline \hline
\# Task & 219,748 & 34,090 & 29,369 \\
\hline
\# Query & 566,967 & 88,021 & 76,159 \\
\hline
Average Task Length & 2.58 & 2.58 & 2.59 \\
\hline
Average Query Length & 2.86 & 2.85 & 2.90 \\
\hline
Average Document Length & 7.27 & 7.29 & 7.08 \\
\hline
Average \# Click per Query & 1.08 & 1.08 & 1.11 \\
\hline
\end{tabular}
\label{table:statistics}
\vspace{-2mm}
\end{table}




\subsection{Dataset and Experimental Setups}
We conduct experiments on the AOL search log data \cite{pass2006picture}. 
Following \cite{sordoni2015hierarchical}, we use the first five weeks as background set, the next six weeks as training set, and the remaining two weeks are divided into half to construct validation and test sets. Note this setting is different from \cite{ahmad2018multitask} that randomly splits search log. 
The background set is used to generate candidate queries for later query suggestion evaluations.
We removed all non-alphanumeric characters from the queries, applied a spelling checker and a word segmentation tool, and lower-cased all the query terms. 

The AOL query log only contains clicked documents under each query and do not record other candidate documents returned to the users. 
Therefore, for a given query, \cite{ahmad2018multitask} aggregated a list of candidate documents, selected from the top documents ranked by BM25 \cite{robertson2009probabilistic} and appended the recorded clicks in the list. 
However, in our preliminary experiments, we observed that many recorded clicks do not have lexical overlap concerning the queries. One possible reason is that we crawled the recorded clicks from the AOL search log in 2017 and many of the clicked documents' content updated since 2006 when the AOL log was recorded.
In such a case, a data-driven model will exploit the differences in lexical overlapping to identify the clicked documents. 
To avoid such a bias in selecting candidate documents, we appeal to the ``pseudo-laebling'' technique, which has been used in prior works \cite{dehghani2017neural} to construct large-scale weekly supervised data to train neural IR models. We first collect the top 1,000 documents for each query retrieved by BM25 and then filtered out the queries, none of whose recorded clicks is in this set of documents. For the resulting queries, we sampled candidate documents from a fixed size window centered at the positions where BM25 ranks the recorded documents. 
Based on this strategy, we sampled 50 candidate documents per query in the test set, and 5 candidates per query for training and validation sets to speed up training and reduce memory requirements.
Besides, following \cite{gao2010clickthrough,huang2013learning,huang2018improving} we only used the document title as its content in our experiments.

\begin{table}[t]
\centering
\caption{Comparison between document ranking models. The paired t-test is conducted by comparing the best and second-best ranking models under each metric, and the test result is presented in bold-faced ($p$-value $<$ 0.05).}
\vspace{-2mm}
\begin{tabular}{l|c|c|c|c|c}
\hline
\multirow{2}{*}{Model} & \multirow{2}{*}{MAP} & \multirow{2}{*}{MRR} & \multicolumn{3}{c}{NDCG}   \\ 
\cline{4-6} 
& & & {@1} & {@3} & {@10} \\ 
\hline \hline
\multicolumn{6}{l}{Traditional IR-models} \\ 
\hline
BM25 \cite{robertson2009probabilistic}  & 0.230 & 0.206 & 0.206 & 0.269 &	0.319 \\ 
QL \cite{ponte1998language} & 0.195 & 0.166 & 0.166 & 0.213 & 0.276 \\
FixInt \cite{shen2005context} & 0.242 & 0.224 & 0.212 & 0.275 & 0.332 \\
\hline \hline
\multicolumn{6}{l}{Single-task Learning} \\ 
\hline
DRMM \cite{guo2016deep} & 0.201 & 0.228 & 0.129 & 0.223 & 0.264 \\
DSSM \cite{huang2013learning}  & 0.283 & 0.307 & 0.188 & 0.231 & 0.341 \\ 
CLSM \cite{shen2014latent}  & 0.313 & 0.341 & 0.205 & 0.252 & 0.373  \\ 
ARC-I \cite{hu2014convolutional} & 0.401 & 0.411 & 0.259 & 0.374 & 0.463 \\
ARC-II \cite{hu2014convolutional}  & 0.455 & 0.465 & 0.309 & 0.434 & 0.521 \\
DUET \cite{mitra2017learning}  & 0.479 & 0.490 & 0.332 & 0.462 & 0.546 \\
Match Tensor \cite{jaech2017match}  & 0.481 & 0.501 & 0.345 & 0.472 & 0.555 \\
\hline \hline
\multicolumn{6}{l}{Multi-task Learning} \\ 
\hline
M-NSRF \cite{ahmad2018multitask}  & 0.491 & 0.502 & 0.348 & 0.474 & 0.557 \\
M-Match Tensor \cite{ahmad2018multitask} & 0.505 & 0.518 & 0.368 & 0.491 & 0.567  \\
CARS & \textbf{0.531} & \textbf{0.542} & \textbf{0.391} & \textbf{0.517}  & \textbf{0.596}  \\
\hline
\end{tabular}
\label{table:ranking}
\vspace{-2mm}
\end{table}

We followed \cite{jones2008beyond} to segment user query logs into tasks. In each user's query sequence $\mathcal{Q}$, we decided the boundaries between tasks based on the similarity between two consecutive queries. 
To this end, we first represented a query by averaging its query terms' pre-trained embedding vectors and computed the cosine similarity between the resulting vectors.\footnote{We used GloVe \cite{pennington2014glove} as the pre-trained word embeddings for this purpose, and used a cosine similarity threshold of 0.5 to segment the tasks.} We discarded the search tasks with less than two queries (no in-task search context). Statistics of our constructed experiment dataset are provided in Table \ref{table:statistics}.

\noindent\textbf{$\bullet$ Evaluation metrics.} 
We used Mean Average Precision (MAP), Mean Reciprocal Rank (MRR), and Normalized Discounted Cumulative Gain (NDCG) as our evaluation metrics for the document ranking task, where we treat the clicked documents as relevant.

For the query suggestion task, we evaluate a model's ability to discriminate and generate the next query.
To test its discrimination ability, we follow \cite{sordoni2015hierarchical} and apply a testing model to rank a list of candidate queries that might follow an anchor query (the second last query of a task). 
We evaluate the rank of the recorded next query among the candidates using MRR.
The candidate queries are selected as the most frequent queries (we consider at most 20 of them) following the anchor query in the background set.
To examine its generation ability, a model is applied to generate the next query and evaluated against the true query based on F1 and BLEU scores \cite{papineni2002bleu}. 
Both scores measure overlapping between the generated query term sequence and ground-truth sequence.

\noindent\textbf{$\bullet$ Baselines.}
We compared CARS with both classical and neural ad-hoc retrieval models.
We consider BM25 \cite{robertson2009probabilistic}, Query likelihood based Language model (QL) \cite{ponte1998language}, and a context-sensitive ranking model FixInt \cite{shen2005context}, as our classical IR baselines for document ranking.
To compare CARS with neural ranking models, we selected the same set of models used in \cite{ahmad2018multitask}, and trained and evaluated them using their publicly available implementations.
To examine CARS's performance in query suggestion, we compared with the sequence to sequence (Seq2seq) approach proposed in \cite{bahdanau2014neural}, an enhanced Seq2seq model with attention mechanism \cite{luong2015effective}, session-based suggestion models HRED-qs \cite{sordoni2015hierarchical}, M-Match Tensor \cite{ahmad2018multitask} and M-NSRF \cite{ahmad2018multitask}.
We used the public implementation of these query suggestion models. 

We carefully tuned the hyper-parameters for the baseline models.\footnote{We tune the hyper-parameters within a range centered around the value (with a window size of 3 or 5) reported in the respective papers.} For all the baselines, we tune the learning rate, dropout ratio, hidden dimension of the recurrent neural network units. For the models involving convolutional neural networks, we tuned the number of filters, and the filter sizes remained unchanged as reported in their original work.

\begin{table}[t]
\centering
\caption{Comparison between query suggestion models. Paired t-test is conducted by comparing the best and second-best models under each metric, and the test result is presented in bold-faced ($p$-value $<$ 0.05).
}
\vspace{-2mm}
\begin{tabular}{l|c|c|c|c|c|c}
\hline
\multirow{2}{*}{Model} & \multirow{2}{*}{MRR} & \multirow{2}{*}{F1} & \multicolumn{4}{c}{BLEU}   \\ 
\cline{4-7} 
& & & {1} & {2} & {3} & {4} \\ 
\hline
\hline
\multicolumn{7}{l}{Single-task Learning} \\ 
\hline
Seq2seq & 0.422 & 0.077 & 8.5 & 0.0 & 0.0 & 0.0 \\ 
Seq2seq + Attn. & 0.596 & 0.555 & 52.5 & 30.7 & 18.8 & 11.4 \\ 
HRED-qs & 0.576 & 0.522 & 48.8 & 26.3 & 15.3 & 9.2 \\
\hline \hline
\multicolumn{7}{l}{Multi-task Learning} \\ 
\hline
M-Match Tensor & 0.551 & 0.458 & 41.5 & 20.6 & 11.5 & 7.0 \\
M-NSRF & 0.582 & 0.522 & 49.7 & 26.7 & 16.0 & 9.9 \\ 
CARS & \textbf{0.614} & \textbf{0.589} & \textbf{55.6} & \textbf{36.2} & \textbf{25.6} & \textbf{19.1} \\
\hline
\end{tabular}
\label{table:suggestion}
\vspace{-2mm}
\end{table}


\noindent\textbf{$\bullet$ Experiment Setup.}
We kept the most frequent $|V| = 80k$ words, and mapped all the others to an \emph{\textless unk\textgreater} token.
We trained CARS end-to-end using mini-batch SGD (with batch size 32) with Adam optimizer \cite{kingma2014adam}.
To stabilize the learning process, we normalized the gradients if their L2 norm exceeds a threshold \cite{pascanu2013difficulty}. 
In CARS, the number of hidden neurons in each of its encoders and decoders were selected from $\{64,128,256\}$.
The initial learning rate and the dropout parameter \cite{srivastava2014dropout} 
were selected from $\{10^{-3},10^{-4}\}$ and $\{0.1,0.2,0.3\}$ based on its performance on validation set, respectively.
We set the hyper-parameters $\lambda_1$, $\lambda_2$, and $\lambda_3$ to $10^{-2}$, $10^{-4}$, and $10^{-1}$ after tuning on the validation set.
We stopped training if the validation performance did not improve for 5 consecutive iterations.
CARS generally stops after 20 epochs of training and each epoch takes 20 minutes on average on a TITAN XP GPU.

\subsection{Experiment Results}

\noindent\textbf{$\bullet$ Evaluation on document ranking.}
We report all models' document ranking performance in Table \ref{table:ranking}.
As we can clearly observe CARS significantly outperformed all the traditional IR and neural IR baselines. Traditional ranking models only focus on keyword matching, which suffer seriously from vocabulary gap.
We group the neural baselines into two groups, single-task learning and multi-task learning models, where the latter can leverage information from the query suggestion task.
All single-task neural ranking models only focus on per-query document matching. Although their learnt query document representations can greatly boosted retrieval performance in every single query, they cannot utilize any search context in a given search task, and therefore only provided sub-optimal search quality.
Comparing with the baseline multi-task learning models, i.e., M-NSRF and M-Match Tensor, which model query formulation chain but not the associated click sequence, CARS complements search context by modeling the past clicks as well and enjoys clear benefit. Later we will perform detailed abalation analysis to decompose the gain into individual components of CARS for more in-depth performance analysis.

\begin{table}[t]
\centering
\caption{Ablation study on CARS.
$\ast$ indicates that the attention in the query recommender (Eq \eqref{eq:attn_sgs}) was turned off to study the impact of search context precisely.}
\vspace{-2mm}
\begin{tabular}{@{}l@{ }|c|c|c|c|c@{}}
\hline
\multirow{2}{*}{CARS Variant} & \multicolumn{3}{c|}{NDCG} & \multicolumn{2}{c}{BLEU} \\
\cline{2-4} \cline{5-6} 
 & {@1} & {@3} & {@10} & {1} & {2} \\ 
\hline
\hline
CARS & 0.391 & 0.517 & 0.596 & 55.6 & 36.2
\\
CARS w/o Attn. & 0.387$^\ast$ & 0.515$^\ast$ & 0.594$^\ast$ & 48.6$^\ast$ & 26.1$^\ast$
\\
\hline 
\hline 
\multicolumn{6}{l}{Ablation on search context} \\ 
\hline
w/o Session Query & 0.379 & 0.505 & 0.586 & 33.7$^\ast$ & 14.2$^\ast$ 
\\
w/o Session Click & 0.356 & 0.485 & 0.568 & 48.2$^\ast$ & 25.6$^\ast$ 
\\ 
\hline
\hline
\multicolumn{6}{l}{Ablation on joint learning} \\ 
\hline
w/o Recommender & 0.379 & 0.505 & 0.585 & - & - 
\\
w/o Ranker & - & - & - & 55.9 & 36.9
\\
\hline
\end{tabular}
\label{table:ablation}
\vspace{-2mm}
\end{table}

\noindent\textbf{$\bullet$ Evaluation on query suggestion.}
We evaluate the models on two bases: a) identifying users' recorded next query from a list of candidate queries (i.e., discrimination ability), and b) generating users' next query (i.e., generation ability).
The comparison results are reported in Table \ref{table:suggestion}.
CARS outperformed all the baselines with significant margins in both of its discrimination and generation abilities.
Although a simple sequence to sequence model only considers consecutive query reformulations rather than the whole task, the attention mechanism still makes it the second best method (i.e., Seq2seq + Attn). This confirms the validity of our constructed local attentive vector (in Eq \eqref{eq:nu}) for query suggestion. CARS improves on it by modeling the entire search task, especially the past click history. Compared with M-Match Tensor and M-NSRF, which model the whole query reformulation chain but still failed to perform in this evaluation, it shows the advantage of our learnt task-level context representation and its utility to the query suggestion task.


\subsection{Abalation Analysis and Discussions}
\label{sec:analysis}
We performed additional experiments by ablating CARS to analyze how and when each component of it adds benefit to the document ranking and query suggestion tasks.
We provide the results of our ablation study in Table \ref{table:ablation} and discuss the significance of them next.

\noindent\textbf{$\bullet$ Benefit of modeling search context.}
To understand the impact of modeling search context, we alternatively striped off the two components from the upper level task embedding layer of CARS (i.e., session-query and session-click encoders). 
First, we turned off the attention between consecutive queries defined in Eq \eqref{eq:attn_sgs} to concentrate on the impact of in-task search context modeling. It slightly affects the model's ranking performance, but generates considerable consequence on the query suggestion quality. This is consistent with our analysis in Table \ref{table:suggestion} and again shows the importance of adjacent queries for query suggestion task.  
As presented in the second block of Table \ref{table:ablation}, without modeling the in-task queries and clicks, CARS loses 3\% and 8.9\% in NDCG@1; and in the meanwhile, it loses 30.7\% and 0.8\% in BLEU-1 (comparing to CARS w/o attention) respectively. 
This result clearly suggests that modeling in-task clicks is more important for the document ranking task and modeling the past queries is crucial for the query suggestion task.

\noindent\textbf{$\bullet$ Multi-task learning v.s. single-task learning.}
We alternatively disabled the document ranker and query recommender components in CARS and reported their performance in the third block of Table \ref{table:ablation}. When the query recommender is disabled, 
the ranking performance of CARS dropped 3.1\% in NDCG@1. This demonstrates the utility of supervision signals from the query recommender to the ranker. However, when the ranker is disabled, the query suggestion performance of CARS was not influenced (and it even became slightly better). We conjecture that since we already encode the clicked documents in the context attentive vector, information from user clicks can be utilized by the model. Therefore, adding training signals from ranker does not provide much additional new knowledge. On the other hand, by training CARS without document ranker, the recommender component can focus more on the query suggestion task, and this might introduce the performance variance. 

%

\begin{figure}[t]
\centering
\vspace{-4mm}
\hspace{-4mm}
\subfloat[Ablation on search context. \label{fig:rsub1}]
{
\includegraphics[width=.50\linewidth]{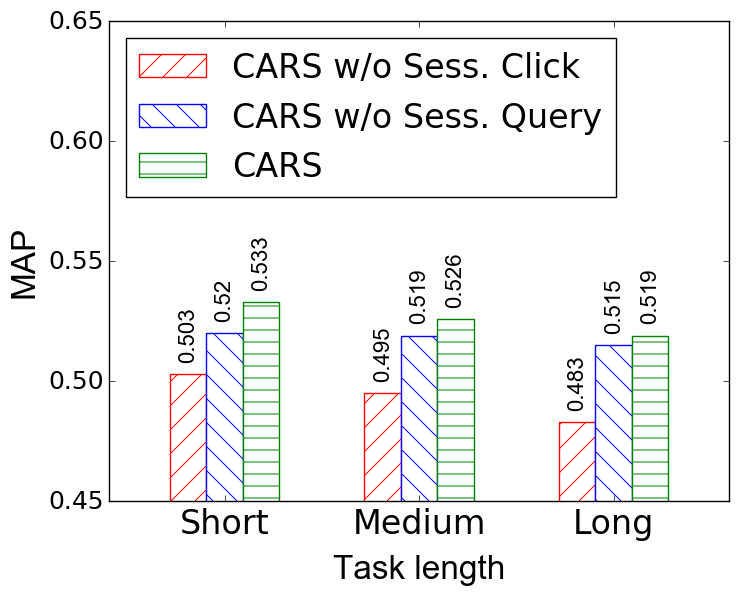}
}
\hfill
\subfloat[Comparing with baselines. \label{fig:rsub2}]
{
\includegraphics[width=.50\linewidth]{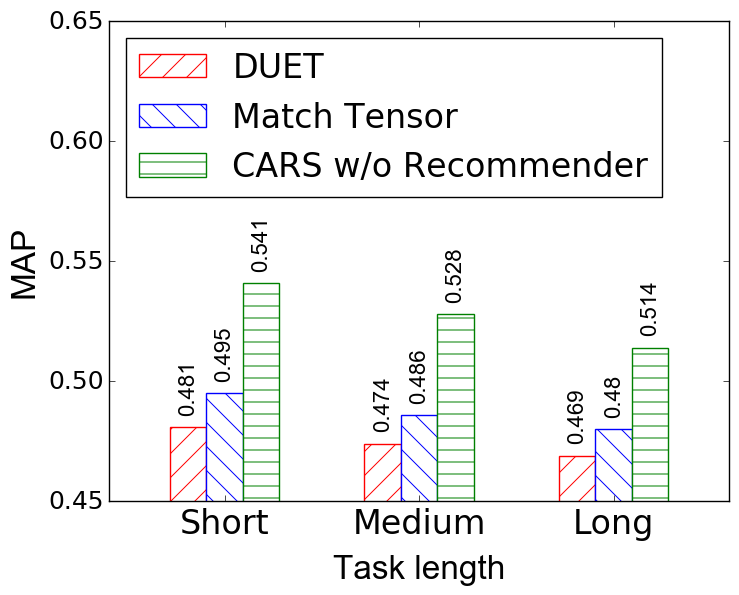}
}

\hspace{-4mm}
\subfloat[Evaluating generation ability. \label{fig:ssub1}]
{
\includegraphics[width=.50\linewidth]{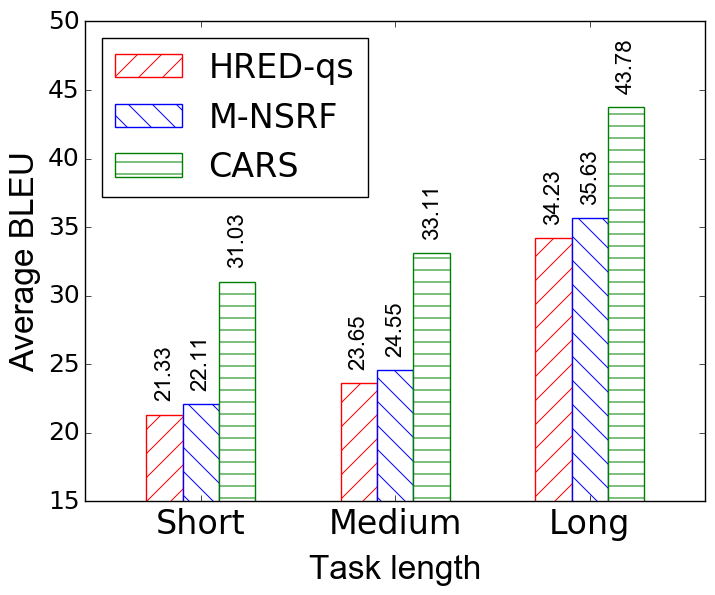}
}
\hfill
\subfloat[Ablation on search context. \label{fig:ssub2}]
{
\includegraphics[width=.50\linewidth]{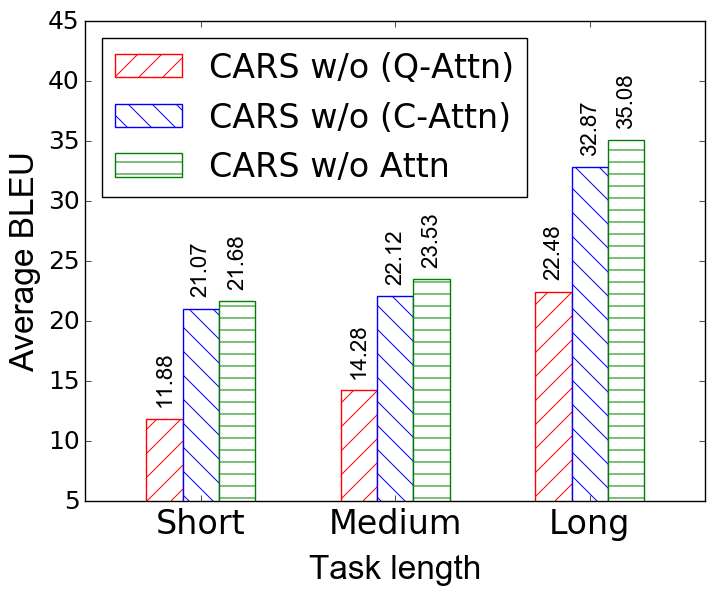}
}
\vspace{-2mm}
\caption{Comparison based on tasks with different lengths.} 
\label{plt:session_length}
\vspace{-4mm}
\end{figure}

\noindent\textbf{$\bullet$ Effect of task length.}
To understand the impact of search context on tasks with different lengths, we performed experiments by splitting the test set into three groups: 
\begin{enumerate}
    \item Short tasks (with 2 queries) -- 66.5\% of the test set
    \item Medium tasks (with 3--4 queries) -- 27.24\% of the test set
    \item Long tasks (with 5+ queries) -- 6.26\% of the test set
\end{enumerate}
As we filtered out queries that do not have any associated clicks when constructing the experiment dataset, we lost some longer tasks; otherwise our test data distribution is similar to \cite{dehghani2017learning}.

We report our findings on the models' document ranking and suggestion performance in Figure \ref{plt:session_length}.
It is clear in Figure \ref{fig:rsub1} that modeling the past in-task clicks is essential for boosting the document ranking performance, especially in long search tasks. MAP dropped 6.9\% and 5.6\% in CARS when session-click encoder was turned off in long and short tasks respectively. However, we can also observe that CARS performed relatively worse in those longer tasks. We hypothesize that those longer tasks are intrinsically more difficult. To verify this, we included two best single-task learning baselines, DUET and Match Tensor, in Figure \ref{fig:rsub2}. And we also turned off query recommender component in CARS to make it focus only on the ranking task. We observed similar trend in those baseline models, i.e., worse performance in longer tasks. In addition, we also found better improvement in the short tasks from CARS to the best baseline rankers than that in the long tasks, 9.3\% v.s., 7.1\%. This indicates modeling the immediate search context is more important.

On the other hand, long tasks amplify the advantage of CARS in the query suggestion task. As we can find in Figure \ref{fig:ssub1}, the query suggestion performance measured by average BLUE score (arithmetic mean among BLUE 1 to 4) of CARS improved 41.4\% from short tasks to long tasks. And compared with the best baseline query recommender that models query reformulation chain in a task, i.e., M-NSRF, better improvement was achieved with short tasks: 40.3\% in short tasks v.s., 22.9\% in long tasks. This further suggests CARS's advantageous sample complexity in learning search context. We also studied the effect of search context modeling with respect to tasks of different lengths in Figure \ref{fig:ssub2}. We turned off the attention between consecutive queries (in Eq \eqref{eq:attn_sgs}) to better illustrate the effect. Clearly, modeling past queries is more important for query suggestion than modeling past clicks; but when the tasks become longer, click history still helps boost the performance.

\begin{figure}
\centering
\vspace{-5mm}
\hspace{-4mm}
\subfloat[\label{fig:complexity1}]
{
\includegraphics[width=.50\linewidth]{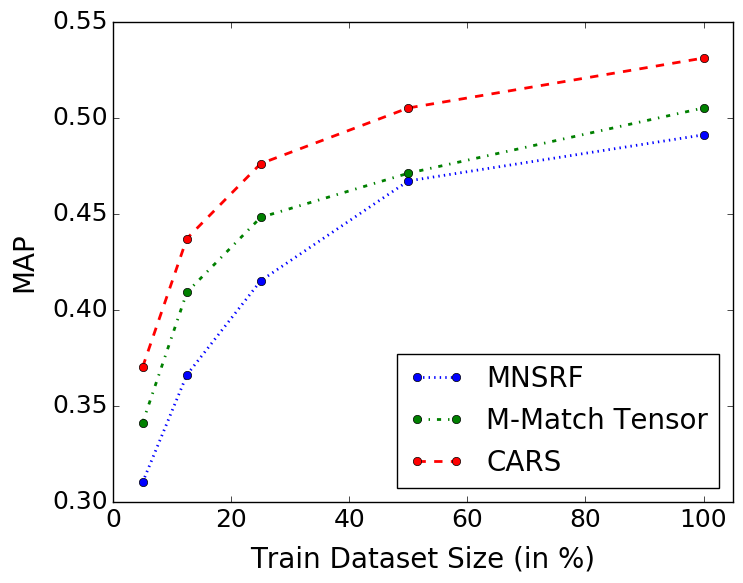}
}
\hfill
\subfloat[\label{fig:complexity2}]
{
\includegraphics[width=.50\linewidth]{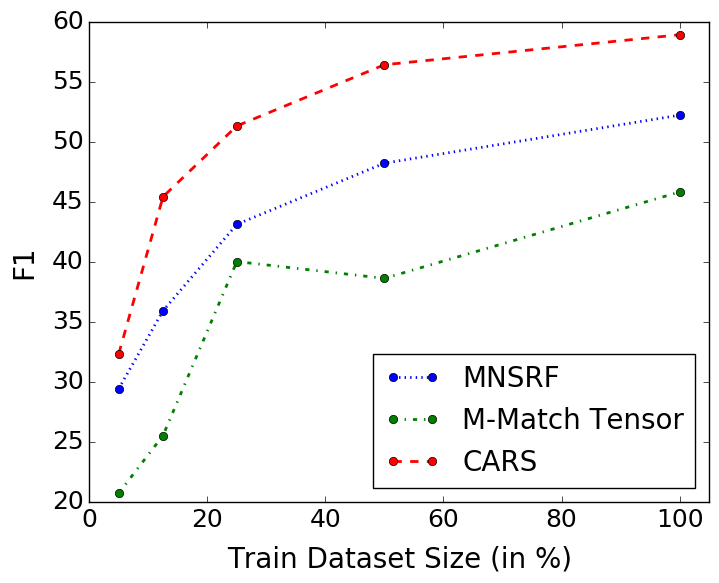}
}
\vspace{-2mm}
\caption{Comparison on sample complexity among the multi-task learning models for (a) document ranking and (b) query suggestion tasks.}
\label{fig:training_size}
\vspace{-2mm}
\end{figure}

\begin{figure}[t]
\centering
\vspace{-2mm}
\subfloat[Click-based attention weight ($\alpha_{j,r}^c$; $\alpha_{j,s}^c$) where $j=1,2$. \label{fig3:ssub2}]
{
\includegraphics[width=.46\linewidth]{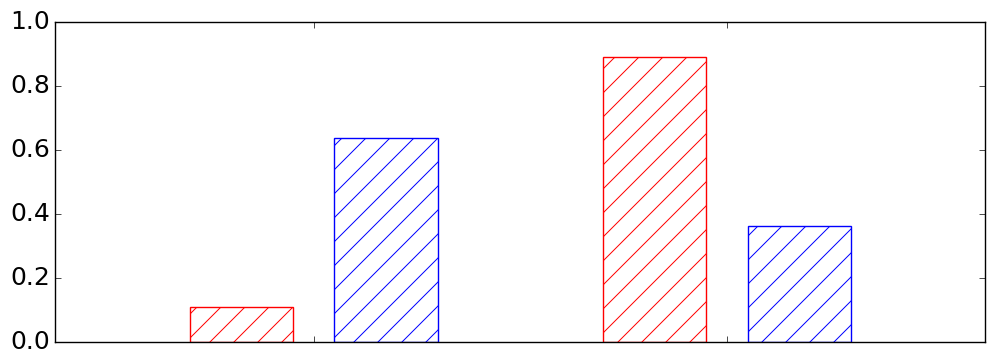}
}
\hfil
\subfloat[Query-based attention weight ($\alpha_{j,r}^q$; $\alpha_{j,s}^q$) where $j=1,2$. \label{fig3:ssub1}]
{
\includegraphics[width=.46\linewidth]{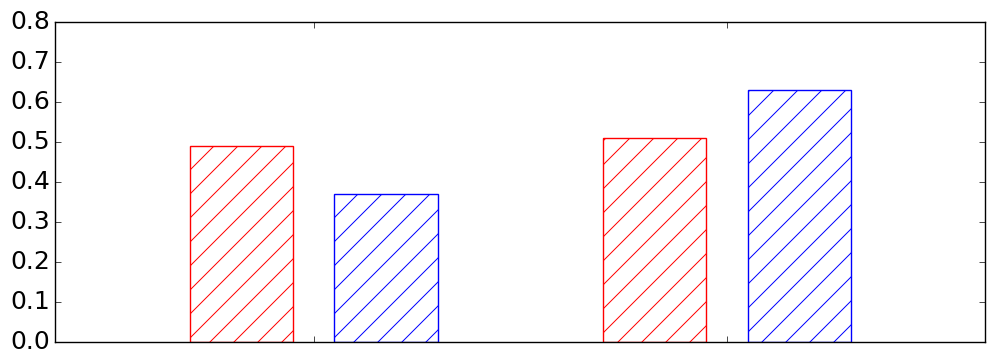}
}
\vspace{-2mm}
\caption{Attention weights ($\alpha_j$ from Eq. \eqref{eq:session_attn}) over session encoder states at Q3 in the search task illustrated in Figure \ref{plt:example}. 
The red and blue bars represent the attention weights for the ranking ($\alpha^{:}_{j,r}$) and suggestion ($\alpha^{:}_{j,s}$) tasks.}
\label{plt:sess_wgt}
\vspace{-4mm}
\end{figure}

\begin{figure*}
\captionsetup[subfigure]{labelformat=empty}
\vspace{-4mm}
\centering
\subfloat[\label{fig2:sub1}]
{
\includegraphics[width=.45\linewidth]{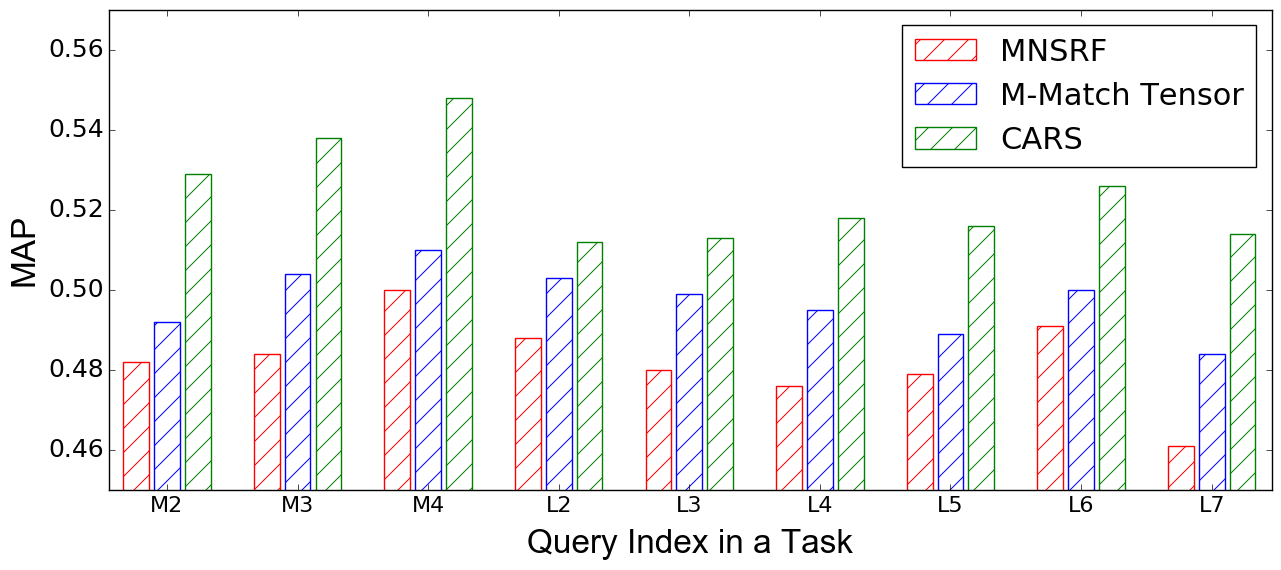}
}
\hfil
\subfloat[\label{fig2:sub2}]
{
\includegraphics[width=.45\linewidth]{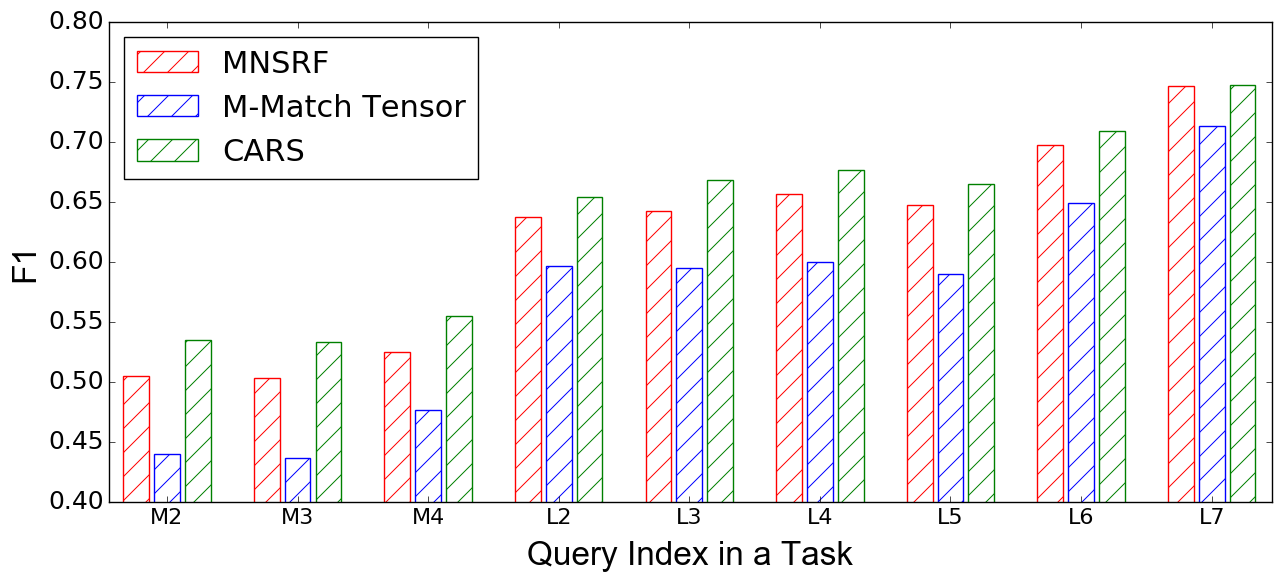}
}
\vspace{-7mm}
\caption{Ranking and suggestion performance comparison between MNSRF, M-Match Tensor, and CARS at different query position in medium (M2--M4) and long (L2--L7) search tasks. The number after ``M'' or ``L'' indicates the query index in a task.}
\label{figure:query_position}
\end{figure*}

\begin{table*}[t]
\centering
\includegraphics[width=0.9\linewidth]{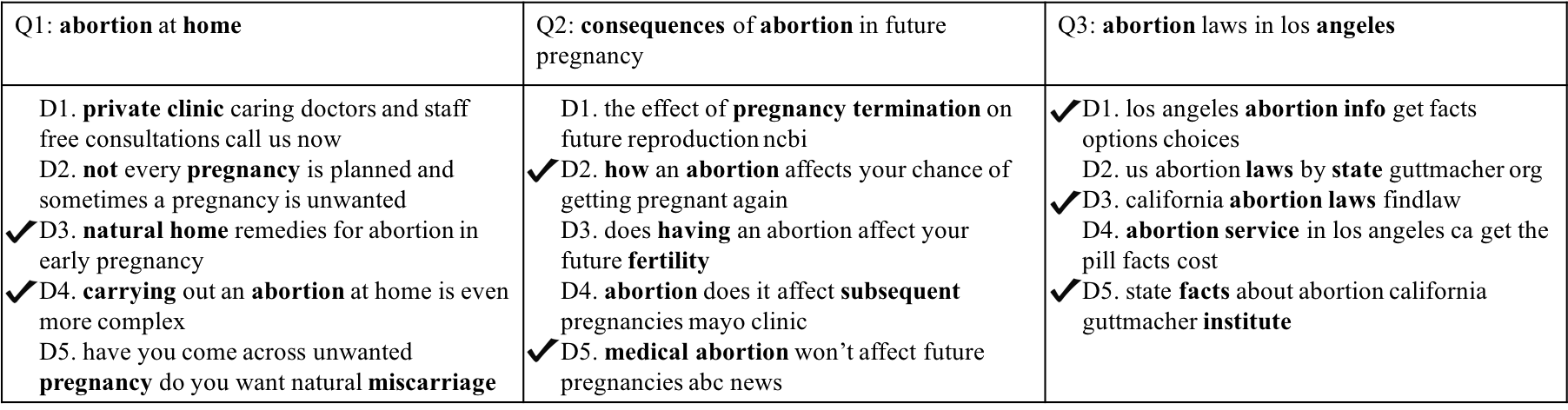}
\vspace{2mm}
\caption{An example search task with three queries and five candidate documents for each query. \text{\cmark} indicates the documents clicked by the user and the words marked in bold are identified as the keywords (gets higher attention weights) by CARS.}
\label{plt:example}
\vspace{-7mm}
\end{table*}



\noindent\textbf{$\bullet$ Performance w.r.t. training data size.}
CARS models both document ranking and query suggestion tasks and consists of multiple encoders and decoders. As a result, it has more than 30 million parameters.\footnote{$W^{gen}$ in query decoder contains about 24 million parameters as the output vocabulary size $|V|$ is 80,000.}
Despite its large number of parameters, CARS converges fairly fast, even with less data, as it effectively exploits training signals from two companion learning tasks.
Figure \ref{fig:training_size} provides a detailed comparison of different models' sample complexity, where we only included the multi-task learning baselines as they are expected to be more effective with less training data.
The fast improving performance of CARS in both  tasks further proves the value of modeling search context and relatedness between the two retrieval tasks in exploiting information buried in users' search activities.

\noindent\textbf{$\bullet$ Effect of modeling task progression.} It is important to study how the modeled search context helps document ranking and query suggestion when a search task is progressing. We compare the performance of CARS with MNSRF and M-Match Tensor at individual query positions in the medium and long search tasks, and report our findings in Figure \ref{figure:query_position}.
It is noticeable that both ranking and query suggestion performance improves steadily as a search task progresses, i.e., more search context becomes available for predicting the next click and query. Both compared baselines benefit from it, especially for document ranking, while CARS improves faster by better exploiting the context.  
One interesting finding is, when the search tasks get longer, the gain of CARS in query suggestion diminishes. As we can observe in Figure \ref{fig2:sub2} that the difference in query suggestion performance between MNSRF and CARS gets smaller from query position L4 to L7. By manually inspecting the test data, we find that users mostly keep submitting the same query when a task gets longer. Moreover, in unusually longer tasks (with more than 7 queries), the user queries are often very short (with only 1 or 2 terms). All the tested models can accurately repeat the previous query by exploiting the context via the attention mechanism.

\noindent\textbf{$\bullet$ Analysis of learnt attention.}
We illustrate a qualitative example in Figure \ref{plt:sess_wgt} and Table \ref{plt:example} to demonstrate the effect of learnt context attention on the document ranking and query suggestion tasks. 
In Table \ref{plt:example}, we highlighted the top two words with the highest self-attention weight in each query and document. Most of them accurately identify the topical focus on the text sequence in both queries and documents. This explains how the learnt representations of query and document help retrieval. In the meanwhile,
Figure \ref{plt:sess_wgt} discloses how the learnt search context representation is leveraged to predict Q3 (i.e., query suggestion) and rank documents for it. To rank the documents under Q3, the clicked documents of Q2 ($\alpha^c_{2,r}=0.91$) impacts more than the other past clicks ($\alpha^c_{1,r}=0.09$); but all the previous in-session queries play an approximately equal role ($\alpha^q_{2,r}=0.51$ and $\alpha^q_{1,r}=0.49$).
On the other hand, to predict Q3 for query suggestion, query Q2 ($\alpha^q_{2,s}=0.63$) impacts more than Q1 ($\alpha^q_{1,s}=0.37$), which is expected. 
And clicks in Q2 ($\alpha^c_{2,r}=0.87$) contributes more than those in Q1 ($\alpha^c_{1,r}=0.13$), which is also meaningful. These results shed light on the potential of using the learnt attention weights for an explanation, e.g., explaining why the documents are ordered in this way based on historical clicks. We leave this as our future work.

  
\section{Conclusion and Future Works}
In this work, we propose a context attentive neural retrieval model for modeling search context in search tasks. It models search context by explicitly utilizing previous queries and clicks from an on-going search task. 
A two-level hierarchical recurrent neural network is introduced to learn search context representations and corresponding dependency structure by jointly optimizing for two companion retrieval tasks, i.e., document ranking and query suggestion. 
Extensive experimentation demonstrates the effectiveness of the proposed search context modeling approach, especially the value of each introduced components to the tasks of document ranking and query suggestion. 

Our work opens up many interesting future directions. 
First, our current solution independently models users' search tasks. 
As different users might have different and consistent search strategies and behavior patterns, modeling across-task relatedness, e.g., users' long-term search interest, becomes necessary. Second, our solution now passively waits for users' next query and click. 
It would be interesting to study it in an online fashion, e.g., reinforcement learning, where the algorithm projects a user's future search actions and optimizes its output accordingly. 
Last but not least, our solution is not limited to web search, but should be applied to any scenario where a user sequentially interacts with a system. 
We would like to explore its utility in a broader application area in future.


\bibliographystyle{acm-reference-format}
\bibliography{acmart}  
\end{document}